\documentclass[smallextended]{svjour3}
\usepackage{epsfig, graphicx}

\smartqed

\usepackage{latexsym}
\usepackage{amsmath}
\usepackage{amssymb}
\usepackage{setspace}
\usepackage{epsfig}
\usepackage{epstopdf }
\usepackage{graphicx}
\usepackage{times}
\usepackage{color, soul}

\usepackage{pgf,pgfarrows,pgfnodes,pgfautomata,pgfheaps,pgfshade}
\usepackage{tikz}
\usepackage{subfigure}

\AppendGraphicsExtensions{.tif}

\journalname{TBA}

\begin{document}




\title{Geodesic  and  Contour Optimization Using Conformal Mapping}



\author{Ricky Fok         \and
        Aijun An \and
        Xiaogong Wang
}


\institute{Ricky Fok \at
              Department of Computer Science,  York University,
 4700 Keele Street,  Toronto, M3J 1P3, Canada. \\
              \email{ricky@cse.yorku.ca}           
           \and
           Aijun An \at
              Department of Computer Science,  York University,
 4700 Keele Street,  Toronto, M3J 1P3, Canada. \\
 	   \email{ann@cse.yorku.ca}
	  \and
	  Xiaogong Wang \at
	Department of Mathematics and Statistics, York University,
 4700 Keele Street,  Toronto, M3J 1P3, Canada. \\
 	\email{stevenw@mathstat.yorku.ca}
}

\date{Received: date / Accepted: date}

\maketitle

\begin{abstract}
We propose   a novel optimization  algorithm for    continuous functions using geodesics and contours  under conformal mapping.
  The algorithm can find multiple    optima
by  first following  a geodesic curve to a local optimum then traveling to the next search area by following  a contour curve. To improve the efficiency, Newton-Raphson algorithm is also  employed in local search steps.
A proposed jumping mechanism based on realized geodesics enables the algorithm to jump to a nearby region and consequently avoid trapping in local optima.
 Conformal mapping is used to resolve numerical instability associated with solving the classical geodesic equations.
  Geodesic flows  under conformal mapping are constructed numerically by using local quadratic approximation.
 The parameters in the algorithm are adaptively chosen to reflect  local geometric features of the objective function.
 Comparisons with many commonly used optimization algorithms including gradient, trust region, genetic algorithm and global search methods have shown
 that the proposed algorithm outperforms most widely used methods in almost all test cases with only a couple of  exceptions.\end{abstract}




\section{Introduction}

\parindent 20pt

Optimization is an essential process in scientific investigation. There are many effective and efficient methods proposed in the literature, see for example  \cite{Polak1997}, \cite{MN2002}, \cite{BL2004}
and \cite{GS2007}.  We propose a new algorithm that solves  optimization problems from a point of view that is different than most of the major methods in the literature. The proposed method builds   one dimensional paths or curves and travels on it with a constant speed to search for the global optimum.

The main idea of the proposed optimization algorithm is built upon  geodesics  which are a generalization of straight lines in Euclidean space that minimizes the
non-Euclidean distance between two points on a given manifold defined by the objective function.
The paths of optimization are constructed numerically by using a local quadratic approximation since there is no   analytical solution to the geodesic equation in a general manifold. To avoid numerical instabilities in converting ill-conditioned matrices, the geodesics are constructed on a manifold under conformal mapping which preserves  intrinsic geometrical features of the objective function. The algorithm will then follow either the geodesics or contours under conformal mapping to search for the optimum. Each contour curve could provide a bridge to a new search area and the constant speed enforced by the algorithm ensures that the search never stops or be trapped at a stationary point. Even with carefully constructed geodesics, the proposal algorithm can still be trapped within one region. In order to search  another promising region, we also build a jumping mechanism by examining the values of the objective function along the geodesics to detect any potential and hidden influence to the geodesic flow from an nearby optimum.

The algorithm can be further improved by integrating with other search methods. 
For example, one can use a few points along the geodesic as starting points to a Quasi-Newton algorithm to improve computational efficiency. From the Quasi-Newton outputs the algorithm is able to change its parameters adaptively for oscillating and smooth objective functions. We found that the resulting algorithm performs well in both types of functions in moderately high dimensions. Furthermore, we built a stopping criterion for the algorithm using Quasi-Newton methods by setting a threshold on the number of the maximum found within tolerance.

The remaining part of the paper is organized as follows.
In Section \ref{DiffGeo} we give an introductory review on differential geometry. Theoretical properties of the proposed algorithm are
established  in Section \ref{theo}.
In Section \ref{Alg}, we give a general description of the algorithm and we also
describe the method of choosing the parameters adaptively.
Numerical results comparing the proposed algorithm
with the Quasi-Newton, genetic algorithm, wedge trust region methods and the global search function in Matlab's global optimization package are
provided in Section \ref{results}.
Finally, the conclusion is given in Section \ref{conclusion}.
 \section{The Main Idea}
\label{DiffGeo}
We generalize the line search method with geodesics in order to discover multiple maxima on a  manifold   conformally related to $\mathbb{R}^n$.

\subsection{  Geodesics and Geodesic Equations}
  We consider a topological manifold that is a second countable and locally compact Hausdorff space. It is  also connected and completely regular.    Detailed discussions can be found in \cite{Boothby2003} and \cite{Lee2010}.
A Riemannian metric on a smooth and differentiable manifold $M$ is a 2-tensor field ${\cal T}^2(M)$ that is symmetric and positive definite. A Riemannian metric thus determines an inner product on each tangent space $T_p(M)$, which is typically written as $g(U, V)$ for $U,V \in T_p(M)$. For an Euclidean space, the
metric matrix (or just metric henceforth) in component form is the Kronecka delta, i.e $g_{ij}=\delta_{ij}$. The inner product $g(U, V)$ in Euclidean reduces to the dot product, $\sum_{ij} \delta_{ij}U^i V^j$, where the sum is over all dimensions. In the Einstein summation convention, it is understood that repeated indices are summed over and the inner product is expressed as $g_{ij}U^i V^j$.

A geodesic is defined to be the path of minimum length for two given distinct points in a connected manifold. It is simply a straight line in  Euclidean space. In a   non-flat manifold, however, it is a curve and no longer a straight line.
 Let $X^i(t)$ denote the local coordinate for the $i$-th dimension for a parameter $t$ which is a time step in our case.
 The geodesic is then characterized  by  a set of partial differential equations, using the Einstein summation convention:
\begin{equation}
\frac{d^2 X^i(t)}{dt^2} + \Gamma^i_{jk} \frac{d X^j(t)}{dt} \frac{d X^k(t)}{dt}  = 0,
\label{Geodesic}
\end{equation}
where $\Gamma^i_{jk}$ are Christoffel symbols defined to be 
\[
\Gamma^i_{jk} = \frac{1}{2} g^{im}\bigg( \frac{\partial g_{mj}}{\partial x^k} +  \frac{\partial g_{mk}}{\partial x^j}  -  \frac{\partial g_{jk}}{\partial x^m} \bigg),
\]
$g_{ij}$ is the metric and $g^{ij}$ is the inverse metric.

There exists a unique vector field on the tangent bundle of manifold, denoted as $TM$, whose trajectories are of the form $(\gamma(t), \gamma^{'}(t))$ where $\gamma$ is the geodesic.
Geodesics play an important role in General Relativity, see \cite{Foster2006}, where the path of a planet orbiting around a star is the projection of a geodesic of the curved 4-D spacetime geometry around the star onto a 3-D Euclidean space.

\subsection{ Conformal Mapping}

Numerical calculations of the Christoffel symbols involve the inversion of the metric and can be unstable and computationally costly. One strategy to avoid such complications is to calculate the Christoffel symbols in a manifold where the metric is easily inverted and then map the results to the manifold desired. We consider the case where the manifold containing information about the objective function is mapped from $\mathbb{R}^n$, where the metric and the inverse metric is the Kronecka delta. In such a case an analytic expression for the Christoffel symbols is available and the costly matrix inversion is avoided.

Each local neighborhood in the new manifold is holomorphic to $\mathbb{R}^n$.  The resulting metric under the conformal mapping is said to be conformally related to the Euclidean metric
\[
g_{ij} = \Psi(x)^2 \delta_{ij},
\]
where the scale factor $\Psi(x) = e^{\phi(x)}$ and $\phi(x)$ is a real valued objective function. Trivially, the manifolds obtained this way are Riemannian as the metric tensors are both symmetric and positive definite. The existence of such mappings is trivial since we assume the existence of $\phi(x)$ to begin with.


\section{Theoretical Properties of the Geodesic  in a Conformally Flat Metric}
\label{theo}

This section investigates the path of the geodesic by considering the direction of its tangent vector and the jumping mechanism used in the algorithm. Unless otherwise stated, the Einstein summation convention is used on all quantities in component form.

\subsection{The Geodesics under Conformal Mapping}
\label{attractor}
\paragraph{Theorem 3.1} {\bf The level curves and the gradient of the objective function $\phi$ are the attractors of the geodesics on any manifold conformally related to Euclidean space.}

\subsection{Jumping Mechanism}
Occasionally, the geodesic can be confined in the neighborhood of a local maximum. Here we discuss a method to estimate the direction of a neighboring maximum from the local maximum using a geodesic so that a jump can be implemented to restart the geodesic along that direction.

Let $l$ be the length of the geodesic, $\gamma$. We define the jumping direction to be along the vector
\[
\Delta \mathbf{x} := \frac{1}{l} \int_{\gamma} \widehat{\phi}(\mathbf{x}) \mathbf{x} \ d \mathbf{x} - \frac{1}{l} \int_{\gamma} \mathbf{x} \ d \mathbf{x},
\]
where the integral is over the geodesic and $\widehat{\phi}$ being the normalized $\phi$ over $t$, . In practice, this is approximated by the sum over all steps along the geodesic
\[
\Delta \mathbf{x} \simeq \frac{1}{T} \sum_{t=1}^{T} [ \widehat{\phi}(\mathbf{x}_t) \mathbf{x}_t  - \mathbf{x}_t],
\]
where $T$ is the total number of steps and $\mathbf{x}_t \in \gamma$. This is just the difference of the weighted mean and the mean position vectors along the geodesic. Suppose that a neighboring maximum exists and the geodesic is symmetric about a local maximum (as it would usually be the case if the geodesic is trapped, for instance, as in Figure \ref{ST}). Then the weighted mean would be slightly biased towards the neighboring maximum. And so $\Delta \mathbf{x}$ would be pointed towards the neighboring maximum. We used a decreasing jump distance for each jump. This is by no means an accurate estimate of the direction to the next maximum. However it has been proven to be sufficient for our algorithm to discover the global maximum in many objective functions of multiple maxima.

\subsection{Solving the Geodesic Equation with a Quadratic Approximation}
\label{stepsizes}
In this subsection we discuss the quadratic approximation used to solve the geodesic equation iteratively. We give an estimation of the adaptive step sizes to ensure that the approximation is valid.
The geodesic equation is
\begin{equation}
\frac{d^2 x^i(t)}{dt^2} + \Gamma^i_{jk} \frac{d x^j(t)}{dt} \frac{d x^k(t)}{dt}  = 0.
\label{Geq}
\end{equation}
In the neighborhood of $\mathbf{x}_t$, the (discretized) approximation to the solution of the geodesic equation is
\begin{equation}
\label{soln}
\mathbf{x}_{t+1} = \mathbf{x}_t + \mathbf{v}_t \delta t + \mathbf{c}_t (\delta t)^2,
\end{equation}
where $\mathbf{v}_t$ is the unit tangent vector of the geodesic at $\mathbf{x}_t$, $\delta t$ is the step size, and
\begin{equation}
\mathbf{c}_t = \frac{1}{2} \frac{d \mathbf{v}}{dt} = \frac{1}{2}[\nabla \phi(\mathbf{x}_t) - 2(\mathbf{v}_t \cdot \nabla \phi(\mathbf{x}_t)) \mathbf{v}_t].
\end{equation}
The tangent vector is estimated by the (normalized) difference $\mathbf{x}_{t} - \mathbf{x}_{t-1}$ and we set the initial tangent vector to be the gradient, $\mathbf{v}_{t=1} = \nabla \phi(\mathbf{x}_{t=1})$. Note that the quadratic approximation (\ref{soln}) is not valid when there exist a component $i$ such that $O(v_{ti} \delta t) \gg O(c_{ti}(\delta t)^2)$, as the approximation is only accurate when the quadratic term is small. The value of $\delta t$ when the linear term equals to the quadratic term in magnitude is $\delta t = t_C$, where
\[
t_C = \min_i \bigg|\frac{v_{ti}}{c_{ti}}\bigg|.
\]
If the quadratic term has opposite sign to the linear term, then $x^i_{t+1} = x^i_{t}$ when $\delta t = t_C$ for some component $i$. Also, when $\delta t = \frac{1}{2} t_C$, $x^i_{t+1} - x^i_t$ is maximized. Therefore, at every time step, the algorithm chooses a step size of $\delta t = \frac{1}{2} t_C$, if it is not smaller than the specified lower bound on $t$ (to be explained in the algorithm section).

Since the geodesic aligns itself with the gradient (or the level curves) as shown in the Section \ref{attractor}, in both cases, the step sizes are
\[
||\mathbf{x}_{t+1} - \mathbf{x}_t || = \frac{3}{4} \frac{1}{| \nabla \phi| }.
\]
It can be found by substituting $\delta t = \frac{1}{2} t_C$ into Equation \ref{soln} and setting $v_i = 1$ in the component parallel to the gradient (or the level curves). Furthermore, as the geodesic travels towards a maximum following the gradient, it has a linear rate of convergence similar to gradient descent.

\section{Algorithm}

\label{Alg}

The algorithm has two parts. The first is a geodesic guided optimization (GEO) algorithm. It estimates the geodesic using the quadratic approximation for a total of $T$ steps. The step size is adaptive and bounded by the validity of the quadratic approximation. Quasi-Newton (QN) optimization may be performed, using points along the geodesic as the starting points. Figure \ref{cam} shows the estimated geodesic moving through three local maxima. The algorithm returns the location of the maximum and its function value along the geodesic, or the one obtained by QN, whichever is the largest.

\begin{figure}
\caption{Geodesic traversing through multiple local maxima in the search space.}
\label{cam}       
\end{figure}

\subsection{Choosing Parameters}
There are cases where the geodesic fails to reach multiple maxima, for instance, as in Figure \ref{ST}. When the objective function is highly oscillatory, the global maximum is less likely to be found by the geodesic. Furthermore, a lower bound on the step size $\delta t_{LB}$ must be specified as an input parameter to prevent the step size to become impractically small in regions of large gradient, but the choice of an appropriate lower bound for any objective function is difficult (if not impossible) to determine. Intuitively, a large  $\delta t_{LB}$  would allow the geodesic to escape from local fluctuations. On the other hand, it may prevent the geodesic from visiting the global maximum.

The second part of the algorithm, Sequential GEO (SGEO), is developed to overcome these difficulties. Information from the geodesic is obtained and passed to SGEO. It includes an estimate of the direction of a neighboring local maximum, $\widehat{\Delta \mathbf{x}}$, an indicator, $k$, to denote whether the geodesic is trapped in a local maximum, and the average distance between the starting points and the end points of  QN, $\bar{R}$, to determine whether the objective function is oscillatory.
\begin{figure}
\caption{Geodesic trapped in a local maximum.}
\label{ST}       
\end{figure}

SGEO calls GEO sequentially with decreasing  $\delta t_{LB}$ for $N$ times. In each subsequent run,  $\delta t_{LB}$ is reduced by a factor of $\alpha$,  determined by requiring that  $\delta t_{LB}$ in the last run to be 1000 times smaller than that in the first run. The next GEO run starts from a position obtained by translating $\mathbf{x^*}$ along $\widehat{\Delta \mathbf{x}}$, with the magnitude and method of the jump determined by $k$. The two mechanisms described above assist in the escape from local maxima. In the case of oscillatory functions, QN is not performed, allowing for a larger number of GEO runs. The algorithm first assumes a non-oscillatory function, and adaptively adjusts its parameters suitable for an oscillatory function by setting a threshold on $\bar{R}$. Finally, we impose a stopping criterion to improve its computational cost. The technical details are described in the follow subsections.  The only inputs needed are the number of GEO runs, $N$, total number of steps $N_T$, the stopping threshold, $N_{th}$, which only depends on the dimensionality, and the initial $\delta t_{LB}$ which is only dependent on the dimensionality and the size of the search region.

\subsection{The first component, GEO}
GEO estimates the geodesic corresponding to a conformally Euclidean metric with the conformal factor given by the objective function up to $T$ steps and evaluates the objective function $\phi(\mathbf{x}_t)$ at every time step, $t$, along the geodesic.

At each $t$, the normalized tangent vector is used to evaluate $\mathbf{x}_{t+1}$ in Equation \ref{soln}. Since the step size is at most of order $O(1 / \nabla_i \phi)$ as discussed in Section \ref{stepsizes}. The geodesic tends to get trapped in regions of large gradient. To avoid this problem we introduce a lower bound on the step size, $\delta t_{LB}$, so that the step size is $\delta t = \max \{0.5 t_C, \delta t_{LB}\}$. The lower bound also ensures that the geodesic has length of at least $T \delta t$.

Now, consider the case where $t_{LB} \gg t_C$. The trajectory is dominated by the quadratic term $\mathbf{c}_t (\delta t)^2$. But at $t=1$, the tangent vector is the unit gradient and so $\mathbf{c}_{t=1} = - \nabla \phi /2$, the geodesic moves against the gradient. An additional minus sign is introduced to $\mathbf{c}_t$ whenever $t_{LB} > 0.5 t_C$ at $t=1$. A backward geodesic that initially moves against the gradient is also estimated by using the initial condition $\mathbf{v}_{t=1} = - \nabla \phi$.

For both the forward and backward geodesics, Quasi-Newton optimization can be performed at every $T_{QN}$ steps. Whenever $\mathbf{x}_{t+1}$ is outside the search region, the algorithm sets $\mathbf{x}_{t+1}$ to be a random point sampled uniformly within the search region. The following quantities are also evaluated to pass to SGEO,  the mean distance between the Quasi-Newton initial position and the solution $\bar{R}$, the normalized mean of $\phi(\mathbf{x}_t)\mathbf{x}_t - \mathbf{x}_t$ over all $t$, and an integer $k \in \{0 , 1 ,2\}$ which characterizes the degree of locality of the geodesics,
\[
k = \begin{cases}
0 & \mbox{if $\phi^*_t$ is not unique in the forward geodesic within tolerance $\forall t$. }    \\
1 & \mbox{if $\phi^*_t$ is unique in only the forward geodesic within tolerance $\forall t$.} \\
2 & \mbox{if $\phi^*_t$ is unique in both geodesics within tolerance $\forall t$,}
\end{cases}
\]
where $\phi^*_t$ is the larger of $\phi(\mathbf{x}_t)$ and the optimized value with QN. The algorithm returns the maximum objective function value and its position along the geodesics as well as the information needed to pass onto SGEO.

\subsection{Algorithm 1: GEO, Geodesic Guided Optimization}
Matlab's fminunc() function is used for the Quasi-Newton optimization. Its input parameters are set as $MaxIter = 200, Tol_{f} = 0.05, Tol_X = 0.01$, where the last two quantities are the tolerances in $\phi$ and $\mathbf{x}$, respectively.

The set of input parameters for GEO is

\begin{itemize}
\item $T$, the number of time steps along the geodesic,
\item $T_{QN}$, the number of time steps between each QN call.
\item $\mathbf{x}_0$, a vector the initial point of the geodesic,
\item $\phi(\cdot)$, the objective function,
\item $\nabla \phi(\cdot)$, the gradient of the objective function,
\item $\mathbf{L}$, a vector containing the lower bounds of the search space,
\item $\mathbf{U}$, a vector containing the upper bounds of the search space,
\item $\delta t_{LB}$, the lower bound on the step size,
\item $s_{QN}$, boolean variable denoting whether QN is performed.
\end{itemize}

The algorithm is as follows

\begin{enumerate}
\item For $t = 1:T$
	\begin{enumerate}
	\item Calculate $\phi(\mathbf{x}_t)$ and $\nabla \phi(\mathbf{x}_t)$.
	\item If $t = 1$, set $\mathbf{x}_t := \mathbf{x}_0$, $k := 1$ and $\delta \mathbf{x}_t := \nabla \phi(\mathbf{x}_t)$,
	\begin{enumerate}
	 	\item else, set $\delta \mathbf{x}_t := \mathbf{x}_t - \mathbf{x}_{t-1}$.
	\end{enumerate}
	\item If $mod(t, T_{QN}) = 0$ and $s_{QN} = 1$,
	\begin{enumerate}
		\item call $QN(\phi(\cdot), \mathbf{x}_t)$ and obtain $\{\phi^{*(1)}_t, \mathbf{x}^*\}$,
		\item set $R^{(1)}_{m} := ||\mathbf{x}^* - \mathbf{x}_t||$ and set $ m := m+1$,
		\item else set $\phi^{*(1)}_t := \phi(\mathbf{x}_t)$.
	\end{enumerate}
	
	\item Calculate the normalized tangent vector $\mathbf{v}_t := \delta \mathbf{x}_t / ||\delta \mathbf{x}_t||$.
	\item Calculate $\mathbf{c}_t := \frac{1}{2}[\nabla \phi(\mathbf{x}_t) - 2(\mathbf{v}_t \cdot \nabla \phi(\mathbf{x}_t))\mathbf{v}_t]$.
	\item Set $t_C := \min_i |v_{ti}/c_{ti}|$, where $i \in \{1,\ldots, D\}$ denotes the $i$-th component.
	\item Set the step size $\delta t := \max \{\frac{1}{2} t_C, \delta t_{LB}\}$.
	\item If  $\delta t =  \delta t_{LB}$ and $t=1$, change the sign of $\mathbf{c}_t := -\mathbf{c}_t$.
	\item Calculate $x_{t+1} := \mathbf{x}_t +  \mathbf{v}_t \delta t + \mathbf{c}_t (\delta t)^2$.
	\item If $x_{(t+1)i} < L_i$ or $x_{(t+1)i} > U_i$ for any component $i$, sample $\mathbf{x}_{(t+1)}$ from a uniform distribution in $[\mathbf{L}, \mathbf{U}]$.
	\item Calculate $\Delta \mathbf{x}^{(1)} := \frac{1}{T}[\sum_{t = 1}^T (\mathbf{x}_t \widehat{\phi}^{*(1)}_t - \mathbf{x}_t)] $, where $\widehat{\phi}^{*(1)}_t = \phi^{*(1)}_t / \sum \phi^{*(1)}_t$
	\item Set $\phi^{*(1)} = \max \phi_t^{*(1)}$.
	\item If $| \phi^{*(1)} - \phi_t^{*(1)}| < Tol_{f}  \phi^{*(1)}$ for all $t$, then set $k =1$. Else set $k = 0 $.
	\end{enumerate}

\item For the backward geodesic, step (1) is repeated with the following adjustments, 
	\begin{enumerate}
	\item $\delta \mathbf{x}_t$ is defined as $\delta \mathbf{x}_t := -\nabla \phi(\mathbf{x}_t)$, in step (1b),
	\item $\phi^{*(2)}_t := QN(\phi(\cdot), \mathbf{x}_t)$ in step (1c(i)),
	\item $R^{(2)}_{m} := ||\mathbf{x}^* - \mathbf{x}_t||$ in step (1c(iii)), and
	\item omitting step (1h).
	\item $\Delta \mathbf{x}^{(2)} := \frac{1}{T}[\sum_{t = 1}^T (\mathbf{x}_t \widehat{\phi}^{*(2)}_t - \mathbf{x}_t)] $ in step (1k).
	\item Set $\phi^{*(2)} = \max \phi_t^{*(2)}$ in step (1l).
	\item If $| \phi^{*(2)} - \phi^{*(1)}| < Tol_{f}  \phi^{*(2)}$ for all $t$ and $k =1$, set $k = 2  $.
	\end{enumerate}
\item Set $\Delta \mathbf{x} := \frac{1}{2}[\Delta \mathbf{x}^{(1)} + \Delta \mathbf{x}^{(2)}]$.	
\item Return $\phi^* := \max \{\phi^{*(1)}, \phi^{*(2)}\}, \mathbf{x}^* := \{ \mathbf{x} | \phi(\mathbf{x}) = \phi^*\}, \bar{R} := \textrm{mean} \{ \mathbf{R}^{(1)},\mathbf{R}^{(2)}\}$, $\widehat{\Delta \mathbf{x}} := \Delta \mathbf{x} / || \Delta \mathbf{x} ||$ and $k$.
\end{enumerate}

\subsection{The second component, SGEO}
SGEO runs GEO sequentially with different parameters. Let $\mathbf{U}$ and $\mathbf{L}$ be vectors denoting the upper and lower bounds of the search space and let $\Lambda = \min (\mathbf{U} - \mathbf{L})$. This algorithm checks whether the objective function is highly oscillating. We use the following criterion that an oscillatory function must satisfy: $\bar{R} < 0.1 \Lambda \sqrt{D}$ in any the first two GEO calls. The reason for limiting to just the first two GEO runs is that $\delta t_{LB}$ gets smaller after each consecutive runs and it is more likely for $\bar{R}$ to be small even for non-oscillatory functions. In the case of a high dimensional ($D > 10$) oscillatory function, no Quasi-Newton optimization is performed to allow for a higher number of GEO runs. Both of which are crucial in locating the global optimum of highly oscillating functions.

The algorithm uses a procedure similar to annealing to reduce $\delta t_{LB}$ for each GEO run. Initially, $\delta t_{LB}^{(n=0)}$ is set to be $\Lambda \sqrt{D} / 100$, where $n$ denotes the $n$-th GEO run. Then the lower bound on $\delta t$ is lowered such that  $\delta t_{LB}^{n} = \alpha^n \delta t_{LB}^{(n=0)}, \alpha \in (0,1)$. The factor $\alpha$ is chosen such that $\delta t_{LB}^{(n=N)}/ \delta t_{LB}^{(n=0)} = 10^{-3}$.

After each GEO call, the initial value, $\mathbf{x}_0^{(n)}$, for the next GEO call is estimated depending on the value of $k$ passed from GEO. Intuitively, $\Delta \mathbf{x}$ would be a vector pointing roughly towards a neighboring maximum. For $k = 0$, the local geodesic is not trapped,
\[
\mathbf{x}_0^{(n+1)} =  \mathbf{x}^{*(n)} + (\alpha^{n} \Lambda) \widehat{\Delta \mathbf{x}}^{(n)}.
\]
For $k=1$, the forward geodesic is trapped and we set $\mathbf{x}_0^{(n+1)}$ to be further away from $\mathbf{x}_0^{(n)}$,
\[
\mathbf{x}_0^{(n+1)} =  \mathbf{x}^{*(n)} + (\alpha \Lambda) \widehat{\Delta \mathbf{x}}^{(n)}.
\]
Finally for $k=2$, when both the backward and forward geodesics are trapped, the method using $\Delta \mathbf{x}$ becomes ineffective as the objective function has similar values along the geodesics - $\Delta \mathbf{x}^{(n)}$ points in the same direction as $\mathbf{x}^{(n)}$. Therefore we simply set $\mathbf{x}_0^{(n+1)}$ to be a point reflected across the midpoint of the search space from $\mathbf{x}_0^{(n)}$,
\[
\mathbf{x}^{(n+1)}_0 :=  \frac{\mathbf{L} + \mathbf{U}}{2} + \alpha^n \Lambda(\frac{\mathbf{L} + \mathbf{U}}{2}- \mathbf{x}^{(n)}_0).
\]

A stopping criterion is imposed to reduce the computational cost. Let $\Phi^{(n)} = \{\phi^{*(n=1)}, \ldots, \phi^{*(n)} \}$ be a series of maxima found up to the $n$-th GEO run, $\phi^* = \max \Phi^{(n)}$ and $N^*$ be the number of elements in $\Phi^{(n)}$ that are within tolerance of $\phi^*$. The algorithm is stopped if $N^* > N_{th}(D)$, where
\[
N_{th}(D) = \begin{cases}
5 & D < 10 \\
10 &  10 \leq D < 20 \\
20 & 20 \leq D \leq 50.
\end{cases}
\]
The algorithm returns $\phi^*$ and $x^*= \{ \mathbf{x} |\phi( \mathbf{x}) = \phi^*\}$.

\subsection{Algorithm 2: SGEO, Sequential Geodesic Optimization} Let $\Lambda= \min (\mathbf{U} - \mathbf{L})$,  the set of input parameters is

\begin{itemize}
\item $N$, the number of GEO runs,
\item $N_T$, total step number,
\item $\delta t_{LB}^{(n=0)}$, starting lower bound on the step size,
\item $N_{th}(D)$.
\end{itemize}

The algorithm is
\begin{enumerate}
\item  Set $\Lambda := \min (\mathbf{U}-\mathbf{L})$, then set $\{ N, \alpha, N_T, \delta t_{LB}^{(n=0)}, T_{QN},s_{QN}\} = \{ 20, 0.7, 500, \Lambda \sqrt{D} / 100, 10,1\}$.
\item Calculate $T := \lfloor \frac{N_T}{N} \rfloor$ and sample $\mathbf{x}_0^{(n=1)}$ uniformly in $[\mathbf{L}, \mathbf{U}]$.
\item For $n = 1:N$
	\begin{enumerate}
	\item Calculate $\delta t^{(n)}_{LB} := \alpha \delta t_{LB}^{(n-1)}$.
	\item Obtain $\{\phi^{*(n)}, \mathbf{x}^{*(n)}, \bar{R}, \widehat{\Delta \mathbf{x}}^{(n)},k \} $ by calling GEO($\mathbf{x}^{(n)}_0,\delta t^{(n)}_{LB},s_{QN}, T, T_{QN}$).
	\item If $D > 10$, $\bar{R} < 0.1 \Lambda \sqrt{D}$, $n \leq 2$ and $s_{QN} = 1$,
	\begin{enumerate}
		\item set  $\{ N, \alpha, N_T,s_{QN}\} := \{ 400, 0.98, 4000,0\}$ and
		\item break and restart current loop with the parameters in the above step in place of those in step 1.
	\end{enumerate}
	\item If $k = 0$, set $\mathbf{x}^{(n+1)}_0 := \mathbf{x}^{*(n)} + (\alpha^n \Lambda) \widehat{\Delta \mathbf{x}}^{(n)}$. Else if $k = 1$, set $\mathbf{x}^{(n+1)}_0 := \mathbf{x}^{*(n)} + (\alpha \Lambda) \widehat{\Delta \mathbf{x}}^{(n)}$. Else set $\mathbf{x}^{(n+1)}_0 :=  \frac{\mathbf{L} + \mathbf{U}}{2} + \alpha^n \Lambda(\frac{\mathbf{L} + \mathbf{U}}{2}- \mathbf{x}^{(n)}_0)$.
	\item If $n = 1$, set $\phi^* = \phi^{*(n=1)}$. Else set $\phi^* := \max \{ \phi^*, \phi^{*(n)}\}$.
	\item Find $N^*$, the number of instances such that $|\phi^* - \phi^{*(n')}| < Tol_{f} \phi^*$, $n' \in \{1, \ldots, n\}$.
	\item If $N^* \geq N_{th}(D)$, exit loop.
	\end{enumerate}
\item Return $\phi^*$ and $\mathbf{x}^* := \{ \mathbf{x} |\phi( \mathbf{x}) = \phi^*\}$.

\end{enumerate}

\section{Numerical Experiments}
\label{results}
  In this section we compare SGEO with other algorithms on test   functions commonly used in the literature. The objective functions\footnote{These are obtained from http://www.sfu.ca/$\sim$ssurjano/optimization.html. We have used the log of the Hartman and Goldstein-Price functions as the values of these functions vary across five orders of magnitude within the search space.} used have dimensions ranging from 2 to 50. Sixteen of which are reasonably smooth, the other twelve are oscillatory. All calculations are performed in Matlab.

Table \ref{tab:1} shows the number of failures in locating the global maximum for existing methods. A success is defined when the estimated function value is within 5\% of the maximum value. If the maximum value is zero, a success corresponds to finding a function value less than 0.05. As it can be seen in Table \ref{tab:1}, the Global Search (GS) method in Matlab's global optimization toolbox outperforms Quasi-Newton (QN), Trust Region (TR), and Genetic Algorithm (GA).

In Table \ref{tab:2} we justify the use of QN and the jumping mechanism in SGEO. The algorithm without QN fails in high dimensions whereas without jumping the algorithm fails in oscillatory cases.

For finding a global maximum from the chosen test functions, we found that the global search method is the best among all commonly used optimization method. We therefore did an extensive  comparison with the global search method in Table \ref{tab:3}. We found that SGEO can discover the true global maximum with a higher chance than GS in the objective functions tested. Our method is more accurate than the global search in many test functions. There is only one function that our method is not as good as the global search. At the same time, the computational cost represented by the number of function calls and the computational time remains similar in most cases.

\begin{table}[ht]
\caption{Number of failures over 100 runs for Quasi-Newton (QN),  Trust Region (TR), Genetic Algorithm (GA), and Global Search (GS).}
\label{tab:1}       
\end{table}

\begin{table}[ht]
\caption{Number of failures over 50 runs for SGEO, without Quasi-Newton (with $T=200$), and without jumping.}
\label{tab:2}       
\end{table}

\begin{table}[ht]
\caption{The number of failures ($N_{failure}$), computational time, and the number of function calls ($N_{call}$) over 50 runs of SGEO and GS.}
\label{tab:3}       
\end{table}

\section{Conclusion}
\label{conclusion}

A new algorithm is proposed in order to find multiple optima of a continuous objective function. The path constructed by the algorithm follows either a geodesic or a contour line. Conformal mapping and the Newton Raphston algorithm are employed to enhance computational efficiency. A built-in jumping mechanism also directs the proposed algorithm to a more promising search area. A stopping criterion is implemented if the same maximum is found too many times.
We are extending this algorithm to handle optimization in high dimensions with contestation. 

\section{Acknoledgements}
This research is supported by the Discovery Grants and Discovery Accelerator Supplement from Natural Sciences and Engineering Research Council of Canada (NSERC).


\end{document}